\documentclass[sigconf]{acmart}

\usepackage{color}

\definecolor{gray}{rgb}{0.8,0.8,0.8} 
\definecolor{green}{rgb}{0, 0.6, 0} 
\definecolor{orange}{rgb}{1, 0.5, 0} 	
\definecolor{mahogany}{rgb}{0.75, 0.25, 0.0}
\definecolor{purple}{rgb}{0.6, 0, 0.6}
\definecolor{darkgreen}{rgb}{0, 0.4, 0}
\definecolor{lightblue}{rgb}{0, 0.8, 1} 
\definecolor{lightgreen}{rgb}{0.9, 1, 0.9} 
\definecolor{lightred}{rgb}{1, 0.3, 0.3} 
\definecolor{navy}{rgb}{0, 0, 0.4}
\definecolor{highlight}{rgb}{1.0, 0.13, 0.32}
\definecolor{black}{rgb}{0.0, 0.0, 0.0}

\newcommand{\shuchang}[1]{\textcolor{black}{#1}}

\AtBeginDocument{%
  }

\copyrightyear{2026}
\acmYear{2026}
\setcopyright{cc}
\setcctype{by-nc-nd}
\acmConference[UIST Adjunct '26]{The 39th Annual ACM Symposium on User Interface Software and Technology}{November 02--05, 2026}{Detroit, MI, USA}
\acmBooktitle{The 39th Annual ACM Symposium on User Interface Software and Technology (UIST Adjunct '26), November 02--05, 2026, Detroit, MI, USA}
\acmDOI{10.1145/3830397.3842691}
\acmISBN{979-8-4007-2855-6/2026/11}

\usepackage{fontawesome5}

\begin{document}

\title{Cyber-Physical Systems for Accessibility and Ability Augmentation: Bridging Diverse Communities}


\author{Shuchang Xu}
\authornotemark[1]
\affiliation{%
  \institution{MIT Media Lab}
  \city{Cambridge}
  \country{USA}}
\email{shuchang@media.mit.edu}

\author{Riku Arakawa}
\authornote{Equal contribution.}
\affiliation{%
  \institution{Carnegie Mellon University}
  \city{Pittsburgh}
  \country{USA}}
\email{rarakawa@cs.cmu.edu}

\author{Mina Huh}
\authornotemark[1]
\affiliation{%
  \institution{UC Berkeley}
  \state{California}
  \country{USA}}
\email{minahuh@berkeley.edu}

\author{Nandi Zhang}
\authornotemark[1]
\affiliation{%
  \institution{University of Rochester}
  \state{New York}
  \country{USA}}
\email{nandi.zhang@rochester.edu}

\author{Tianyu Zhang}
\authornotemark[1]
\affiliation{%
  \institution{University of Rochester}
  \state{New York}
  \country{USA}}
\email{tzhang71@cs.rochester.edu}

\author{Wazeer Zulfikar}
\authornotemark[1]
\affiliation{%
  \institution{MIT Media Lab}
  \city{Cambridge}
  \country{USA}}
\email{wazeer@media.mit.edu}

\author{Ruei-Che Chang}
\authornotemark[1]
\affiliation{%
  \institution{University of Michigan}
  \city{Ann Arbor}
  \country{USA}}
\email{rueiche@umich.edu}

\author{Yotam Sechayk}
\authornotemark[1]
\affiliation{%
  \institution{University of Tokyo}
  \city{Tokyo}
  \country{Japan}}
\email{ysechayk@acm.org}

\author{Huamin Qu}
\affiliation{%
  \institution{The Hong Kong University of Science and Technology}
  \city{Hong Kong}
  \country{China}}
\email{huamin@cse.ust.hk}

\author{Amy Pavel}
\affiliation{%
  \institution{UC Berkeley}
  \state{California}
  \country{USA}}
\email{amypavel@eecs.berkeley.edu}

\author{Franklin Mingzhe Li}
\affiliation{%
  \institution{UNC-Chapel Hill}
  \city{North Carolina}
  \country{USA}}
\email{fmli@unc.edu}

\author{Yukang Yan}
\affiliation{%
  \institution{University of Rochester}
  \state{New York}
  \country{USA}}
\email{yukang.yan@rochester.edu}

\author{Brian A. Smith}
\affiliation{%
  \institution{Columbia University}
  \city{New York}
  \country{USA}}
\email{brian@cs.columbia.edu}

\author{Pattie Maes}
\affiliation{%
  \institution{MIT Media Lab}
  \city{Cambridge}
  \country{USA}}
\email{pattie@media.mit.edu}

\renewcommand{\shortauthors}{Xu et al.}

\begin{abstract}
The powerful convergence of wearables, robotics, extended reality, and smart environments is expanding the design space for cyber-physical systems (CPS) that support and augment human abilities in daily life. 
By sensing real-world contexts, modeling user needs, and providing situated assistance, these systems can improve accessibility for people with disabilities while enhancing broader human abilities such as perception, memory, learning, and mobility. 
However, realizing this potential requires addressing key challenges in context sensing, user modeling, adaptive interaction, privacy, and evaluation to ensure that CPS are reliable and effective in real-world contexts. 
This workshop will bring together researchers and practitioners across HCI, AI, wearables, robotics, XR, smart environments, accessibility, and ability augmentation to examine shared strategies and challenges for designing accessibility- and ability-centered CPS. 
Through panel discussions, interactive demos, and mixed-group design activities, participants will identify recurring design principles, technical challenges, and future directions for CPS that support and augment human abilities in real-world settings. For details, please visit: \faIcon[regular]{compass}\ \url{https://cps4all.github.io}.
\end{abstract}


\sloppy

\maketitle

\section{Introduction}
Recent advances in wearables \cite{harrison2011omnitouch, tao2021altering}, robotics \cite{kari2025reality, suzuki2019shapebots}, extended reality \cite{cheng2021semanticadapt, evangelista2022auit}, and smart environments \cite{suzuki2020roomshift, ahmetovic2016navcog} are expanding the design space for Cyber-Physical Systems (CPS) that support and augment human abilities in everyday life. 
By sensing real-world contexts, modeling user needs, and delivering situated assistance in physical environments, these systems create new opportunities to enhance human capabilities such as perception \cite{chang2024worldscribe, xu2025branch, jain2023front}, memory \cite{zulfikar2024memoro, xu2024memory, maniar2025mempal,xu2026remiassist}, learning \cite{DBLP:conf/chi/ArakawaYK22, gunturu2024augmented, das2026co, huh2025vid2coach}, and mobility \cite{anderson2013youmove, xiong2024reach, wang2026wearable}. 
They can also address long-standing accessibility barriers for people with disabilities across domains such as independent living \cite{lee2024cookar, li2021non, arakawa2024prism, huh2025vid2coach, ning2025aroma,huh2026designing}, navigation \cite{ahmetovic2016navcog, jain2024streetnav, xu2020virtual}, and social participation \cite{teng2024help, xu2025danmua11y, bandukda2024context}.


\shuchang{Although CPS research spans diverse technologies and user communities, many systems face shared challenges in context sensing \cite{DBLP:journals/imwut/ArakawaYMNRDRMC22, arakawa2024prism, chang2024worldscribe}, user modeling \cite{danry2026mind, shaikh2025creating}, adaptive interaction \cite{yan2018eyes, lai2025adaptique,DBLP:conf/uist/ArakawaPPLG25}, privacy \cite{zhang2024designing,DBLP:journals/imwut/PatidarAGMSVTSSMGA25}, and long-term evaluation \cite{long2025facilitating}. While prior work at UIST has made substantial advances and offered diverse solutions to individual challenges, there have been limited opportunities for researchers across these areas to exchange insights, compare approaches, and develop a shared agenda for CPS that advances accessibility and ability augmentation in real-world settings.}

To bring together research communities around these emerging challenges, this workshop proposal builds on our successful ``Accessible Cyber-Physical Activities'' workshop~\cite{DBLP:conf/uist/ArakawaLZHP0CY25} at UIST 2025, which convened 45 participants from diverse backgrounds and demonstrated strong interest in broader questions of human augmentation. 
The proposed workshop extends, rather than repeats, the prior effort in three key ways. 
First, it broadens the conversation from accessibility-centered CPS to ability augmentation across technologies, applications, and research communities. 
Second, it introduces more hands-on activities, including interactive demos and mixed-group design activities. 
Third, it places stronger emphasis on real-world deployment and evaluation by connecting academic researchers with industry practitioners and end users.

This workshop will convene researchers and practitioners from HCI, AI, sensing, ubiquitous computing, wearables, robotics, AR/VR, smart environments, accessibility, and human augmentation to exchange transferable insights and shape future directions for CPS research. 
Drawing on methods for fostering interdisciplinary exchange~\cite{bennett2012collaboration,hubbs2020toolbox}, the workshop will feature keynotes, panel discussions, interactive demonstrations, and mixed-group design activities. These activities will help participants connect perspectives across domains, identify shared challenges, and develop a research agenda for CPS that augment human abilities in everyday life. 
The organizing committee will synthesize key workshop outcomes into a public report and maintain a participant communication channel to support continued exchange beyond the event.

\begin{figure}
    \centering
    \includegraphics[width=0.81\linewidth]{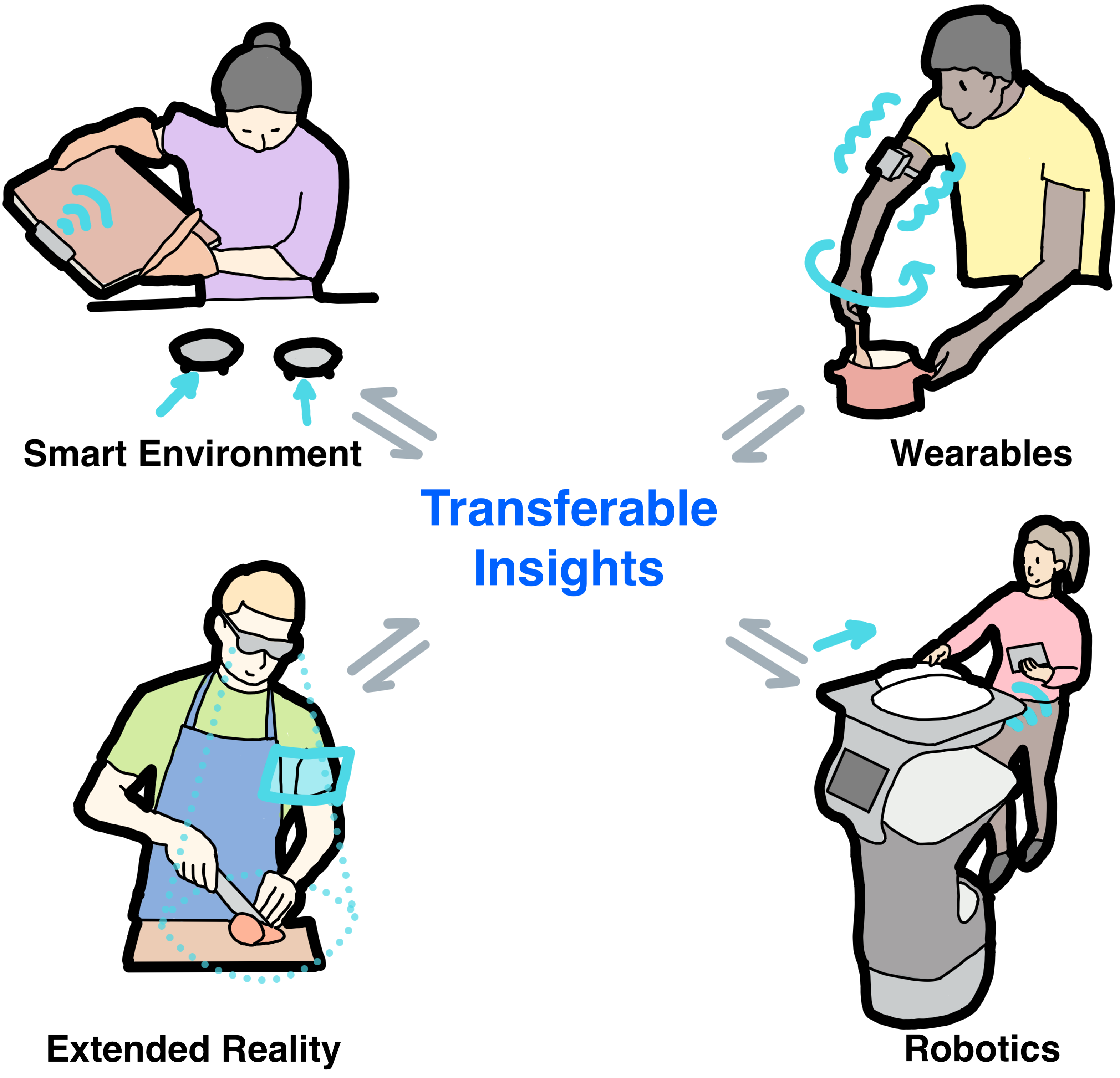}
    \caption{Cyber-physical systems from different fields (smart environments \cite{han2025towards}, wearables \cite{nith2024splitbody}, extended reality \cite{huh2025vid2coach,ning2025aroma}, robotics \cite{kari2025reality}) enhance human capabilities in everyday activities
    . This workshop will bring together these diverse communities to exchange transferable insights.}
    \label{fig:Fragmented_Research}
\end{figure}

\section{Topics}\label{sec}
\shuchang{The workshop will examine five key challenges in designing CPS for accessibility and ability augmentation in real-world settings:}

\begin{enumerate}
\item \textbf{Sensing and Modeling:} How can CPS robustly sense, interpret, and model users across diverse real-world contexts?
\item \textbf{Interaction Paradigms:} How can CPS provide situated assistance while preserving user agency during daily tasks?
\item \textbf{End-User Adaptation:} How can CPS adapt to diverse user needs, abilities, preferences, and physical environments?
\item \textbf{Real-World Evaluation:} How should CPS be evaluated to assess real-world usability and long-term impact?
\item \textbf{Privacy and Ethics:} How can CPS protect privacy, safety, and bystander consent in everyday physical spaces?
\end{enumerate}

\shuchang{These topics will help participants identify recurring design strategies, technical challenges, and future directions for CPS that responsibly augment human abilities in everyday life.}

\section{Workshop Plan}
\label{sec:plan}
The workshop will foster meaningful interdisciplinary exchange by combining lightweight preparation, invited talks, panel discussions, demos and posters, and mixed-group design activities. 

\subsection{Pre-Workshop Preparation}
Before the workshop, participants will complete a short survey describing their background, goals, relevant expertise, and whether they plan to share a demo or poster. With consent, responses will be summarized on the workshop website to help attendees learn about one another in advance. Organizers will also use the responses to form diverse discussion groups of 4--6 participants, balancing technical areas, application domains, and user perspectives.

\subsection{Workshop Activities}

\subsubsection{Opening and Icebreaker (30 minutes).} The workshop will begin with an overview of the agenda and goals, followed by an icebreaker in the pre-assigned groups. Participants will introduce themselves and identify shared interests, complementary expertise, and questions to explore during the workshop.

\subsubsection{Keynotes (90 minutes).} 
We will invite two experts in CPS and human augmentation, one from academia and one from industry, to deliver keynotes. Each speaker will give a 30-minute talk followed by a 15-minute Q\&A. 
The keynotes will highlight recent advances in the field and set the stage for subsequent discussions.

\subsubsection{Panel Discussion (45 minutes).} We will invite 5--6 panelists with diverse perspectives across CPS technologies, application domains, and user communities to discuss open challenges and emerging opportunities in the field. Attendees will be invited to engage directly with the panelists during the discussion.

\subsubsection{Demo \& Poster Session (60 minutes).} Participants who indicate interest in the pre-workshop survey will share demos, prototypes, videos, or posters at tables and stands. This format will allow attendees to move between projects, compare approaches, exchange insights, and identify potential collaborators.

\subsubsection{Mixed-Group Design \& Discussion (120 minutes).} During the first half of the session, participants will return to their pre-assigned groups for scenario-based design activities. 
\shuchang{
Each group will select a specific user scenario and address the key questions outlined in Section~\ref{sec}.} 
Guided prompts will help groups identify shared technical challenges, design strategies, evaluation barriers, and opportunities for collaboration. 
Organizers will rotate among groups to keep discussions focused, inclusive, and productive. 
During the second half of the session, all groups will reconvene to share their design outcomes and contribute their main insights to a shared affinity-mapping activity. The resulting themes will form the basis of a public workshop report.

\subsubsection{Closing Session (20 minutes).} The workshop will conclude with a synthesis of key takeaways and next steps. We will also open a shared communication channel and repository for workshop artifacts to support continued collaboration after UIST.

\subsection{Post-Workshop Activities}
After the workshop, we will: 
(1) publish a summary of key insights and discussion highlights on the workshop website; 
(2) share workshop artifacts and presentation slides with participants; 
(3) invite attendees to join a Discord server for continued sharing and collaboration; and 
(4) summarize major outcomes in a public workshop report on arXiv to broaden the workshop’s impact.







\section{Organizers}
The organizers bring relevant and diverse expertise aligned with the workshop theme, including wearables \cite{DBLP:conf/uist/ArakawaPPLG25, DBLP:conf/uist/PrISM-Observer,DBLP:journals/imwut/ArakawaYMNRDRMC22,li2019fmt,wang2026wearable,xu2020virtual}, robotics \cite{huh2026designing,lyu2025signaling,han2024co}, extended reality \cite{zhang2025following, lu2026capability,gunturu2025realitysummary}, smart environments \cite{DBLP:journals/imwut/PatidarAGMSVTSSMGA25,maes2005attentive,jain2024streetnav}, accessibility \cite{huh2023genassist, sechayk2025veasyguide, chang2024worldscribe, ning2025aroma, xu2025branch,xu2026sonic}, and ability augmentation \cite{zulfikar2024memoro, smith2018rad, xu2024memory,fang2025leveraging}. Several organizers have successfully organized related workshops at UIST \cite{DBLP:conf/uist/ArakawaLZHP0CY25} and CHI \cite{DBLP:conf/chi/GlazkoHJPM25,danry2025human}. One organizer also brings lived experience with accessibility needs, helping ensure that the workshop is grounded in community perspectives.

\section{Call for Participation}

We welcome participants whose work explores CPS for accessibility and human ability augmentation. Relevant areas include, but are not limited to, HCI, AI, sensing, ubiquitous computing, wearables, robotics, AR/VR, smart environments, accessibility, human augmentation, privacy and ethics, and real-world evaluation.

\section{Accessibility of the Workshop} 
We strive to make the workshop accessible and welcoming to participants from diverse communities and with diverse abilities. The pre-workshop survey will invite participants to share any accessibility needs, and we will work to accommodate these needs.



\bibliographystyle{ACM-Reference-Format}
\bibliography{references}

\appendix

\end{document}